\documentclass[aps,prd,twocolumn,a4paper,superscriptaddress,10pt,floatfix]{revtex4-1}
\usepackage{graphicx}
\usepackage{multirow}
\usepackage{latexsym}
\usepackage{hyperref}
\usepackage[british]{babel}
\usepackage{amsmath}
\usepackage{color}
\usepackage[dvipsnames]{xcolor}

\begin{document}
\newcommand{\be}{\begin{equation}}
\newcommand{\ee}{\end{equation}}
\newcommand{\bq}{\begin{eqnarray}}
\newcommand{\eq}{\end{eqnarray}}

\title{Updated constraints on Regge-Teitelboim gravity}
\author{S. R. Pinto}
\email{up202004386@edu.fc.up.pt}
\affiliation{Centro de Astrof\'{\i}sica da Universidade do Porto, Rua das Estrelas, 4150-762 Porto, Portugal}
\affiliation{Faculdade de Ci\^encias, Universidade do Porto, Rua do Campo Alegre, 4150-007 Porto, Portugal}
\author{C. J. A. P. Martins}
\email{Carlos.Martins@astro.up.pt}
\affiliation{Centro de Astrof\'{\i}sica da Universidade do Porto, Rua das Estrelas, 4150-762 Porto, Portugal}
\affiliation{Instituto de Astrof\'{\i}sica e Ci\^encias do Espa\c co, Universidade do Porto, Rua das Estrelas, 4150-762 Porto, Portugal}
\date{\today}

\begin{abstract}
In the Regge-Teitelboim model, gravity is described by embedding the space-time manifold in a (usually flat) fixed higher-dimensional background, where the embedding coordinates, rather than the metric tensor, are the dynamical degrees of freedom. Stern \& Xu extended the Regge-Teitelboim framework to encompass scenarios where the background embedding space is not flat, noting that when the background is a five-dimensional de Sitter space, the Robertson-Walker manifold undergoes a transition from a decelerating phase to an accelerating one. Previously, we constrained this model using only low-redshift observations. Here we further explore the observational constraints on this scenario, and report significantly more stringent constraints by including high-redshift data, specifically from the cosmic microwave background. Our results are consistent with $\Lambda$CDM, with the putative model-specific energy component responsible  for the recent acceleration being constrained to $\Omega_{RT}<0.006$ and the de Sitter curvature radius in units of the Hubble constant being constrained to $LH_0>1.45$, both at the 95 percent confidence level.
\end{abstract}
\maketitle

\section{\label{introd}Introduction}

Identifying the physical mechanism driving the observed low-redshift acceleration of the universe is the most compelling goal of modern fundamental cosmology. The $\Lambda$CDM model, which has the minimal number of additional parameters, is broadly consistent with all existing data (despite some observational hints of possible inconsistencies, colloquially dubbed 'tensions'). Nevertheless, the observed value of the cosmological constant defies contemporary theoretical predictions and, if proven correct, it would require fine-tuning or some other radical departure from current knowledge. One such radical solution to explain the acceleration is by altering the behavior of gravity itself, but that will have specific observational fingerprints, which one can look for in the ever-improving available data.

In a recent work \cite{artigo1} we have used low-redshift background cosmology data to place quantitative constraints on three separate modified gravity models, each of which aims to explain the low-redshift acceleration through a different non-standard physical mechanism. Among these three, we found the Regge-Teitelboim model \cite{Regge} to be the most interesting one. In this class of models, the conceptual premise is that gravity is described by embedding our usual space-time manifold in a fixed higher-dimensional background, and to regard the embedding coordinates as dynamical degrees of freedom. These embeddings lead to additional source terms in the Einstein equations which, in certain circumstances, could lead to accelerating solutions. While most of these embeddings are easily ruled out by low-redshift data, we found that the one proposed by Stern \& Xu \cite{Stern} was consistent with this data, at the cost of requiring a non-standard value of the matter density. One difference between this choice of embedding and others previously proposed in the literature is that in this case the higher-dimensional background is non-flat. More specifically, acceleration would allegedly arise in a 5-dimensional de Sitter space ($dS_5$). 

In the present work we extend our earlier work by studying the full cosmological field evolution of the Stern-Xu class of Regge-Teitelboim gravity and using a standard Bayesian likelihood analysis \cite{Verde} for confronting it with the latest datasets. As one might expect, the main difference in constraining power (as compared to our earlier work) is provided by the addition of cosmic microwave background (CMB) data. As we will see, under the assumption that the model is a parametric extension of $\Lambda$CDM (i.e., if a cosmological constant is still allowed to be non-zero, despite the presence of an alternative mechanism for acceleration), going beyond the low-redshift background cosmology data and considering a broad set of currently available cosmological data constrains the model to be effectively indistinguishable from $\Lambda$CDM.

\section{The Stern-Xu class of Regge-Teitelboim models}

Regge-Teitelboim (RT) gravity \cite{Regge} is an alternative to the canonical theoretical framework of classical general relativity, suggesting that the origin of the dynamics of gravity is in the embeddings of the usual (3+1)-dimensional  space-time within a fixed higher-dimensional background, rather than in the metric tensor. Unlike conventional approaches, where solutions to Einstein's equations rely solely on the energy-momentum tensor, RT gravity has additional source terms arising from the embedding of the space-time. These extra terms come from considering that rather than the metric tensor, the embedding coordinates are the dynamical degrees of freedom. In this section we provide a brief overview of the model's salient features, as well as of our earlier results \cite{artigo1}. In passing, we should caution the reader that the RT approach has been criticized by other authors \cite{Deser}.

The above paradigm does not prescribe a specific choice of embedding, which is precisely the reason for the criticism in \cite{Deser}. The author argues that the physical predictions of the model can change based on arbitrary gauge choices made in the embedding, which undermines its physical viability. Furthermore, \cite{Deser} points out the difficulty of recovering the familiar linearized weak-field approximation, $g=\eta+h$, directly in terms of the embedding coordinates. While this latter issue is often not directly addressed in subsequent works on RT gravity, it is important to note that, rather than causing unphysical effects, different embedding choices are often interpreted as defining distinct classes of cosmological models, each with its own characteristic cosmological behavior. In principle, such individual characteristics could lead to observational tests of these different classes.

From a practical point of view, the most important consequence of the embedding procedure is that it will lead to additional source terms in the standard Einstein equations, and one may therefore consider the possibility that these additional terms are responsible for the recent acceleration of the universe. From this perspective, we consider these models to be purely phenomenological, as they are not derived from a unique, underlying fundamental theory but are instead constructed to reproduce observed cosmological phenomena. Their primary aim is to fit the data--in this case, to model the recent acceleration of the universe--rather than to explain the ultimate origin of the additional terms or behaviors they introduce. Note that in principle, in these models, the cosmological constant can be assumed to vanish, in which case the models will not have a $\Lambda$CDM limit. Nevertheless, one can also allow for its presence, which will make them parametric extensions of $\Lambda$CDM. This has already been done in \cite{artigo1} and will also be done in what follows.

Earlier studies of RT gravity considered flat five-dimensional embeddings \cite{Davidson,Fabi}, but this assumption was relaxed by Stern \& Xu \cite{Stern}, who have explored non-flat cases. Specifically, these authors claim that by embedding the Robertson-Walker manifold within curved backgrounds such as $R^{4,1}$, $AdS_5$, and $dS_5$, the formalism of RT gravity extends naturally to more realistic cosmological scenarios. Notably, they pointed out that an embedding in five-dimensional de Sitter $dS_5$ space can lead to late-time accelerating solutions. In this case, the evolution is determined by two free parameters: the curvature of the background de Sitter space (we will denote the corresponding curvature radius by $L$) and and effective dimensionless density, quantifying the contribution of the RT source term to the Friedmann equation, which in \cite{artigo1} was denoted $\Omega_c$. We retain this definition in the present section, but note that in Sect. \ref{extens} a related and numerically more convenient but different definition will be introduced.

Defining for convenience the dimensionless Hubble parameter $E(z)=H(z)/H_0$, one can show \cite{artigo1} that the Friedmann equation can be written
\be
E^2(z)=\Omega_m(1+z)^3+\frac{\Omega_c(1+z)^4}{\sqrt{(LH_0)^2E^2-1}}\,,
\ee
where $\Omega_m$ is the standard matter density and there is an additional relation between the model parameters
\be
(LH_0)^2=1+\left(\frac{\Omega_c}{1-\Omega_m}\right)^2\,.
\ee
Note that in the limit ($\Omega_c\to0$, $L\to\infty$) one recovers the Einstein--de Sitter model, with $\Omega_m=1$.

\subsection{Previous low-redshift constraints}

Previous work, using low-redshift data from the Pantheon catalog \cite{Scolnic,Riess} and the compilation of Hubble parameter measurements reported in Farooq \textit{et al.} \cite{Farooq}, and analytically marginalizing the Hubble constant \cite{Homarg}, derived the following constraints \cite{artigo1}
\bq
\Omega_c &=& 0.30 \pm 0.03\\
\Omega_m &<& 0.06\,,
\eq
the former being a one-sigma constraint and the latter a two-sigma upper limit. These would lead to a best fit dimensionless curvature radius $LH_0\sim1.04$. The two densities are anti-correlated, and if one were to (purely phenomenologically) allow for the unphysical possibility of a negative matter density the peak of the likelihood would shift to $\Omega_m\approx-0.13$ and $\Omega_c\approx +0.41$, with little change in the preferred dimensionless curvature radius. One additional statistical cost of allowing for negative matter densities is that the model would overfit the data, which was not the case if only non-negative matter densities are allowed- Seemingly, in this scenario the two parameters, $\Omega_m$ and $\Omega_c$, add up to the standard value of the matter density, but the data prefers a dominant contribution from $\Omega_c$.

Allowing for the presence of a cosmological constant, along with the higher-dimensional term, is straightforward. The Friedmann equation simply becomes
\be
E^2(z)=\Omega_m(1+z)^3+\Omega_\Lambda+\frac{\Omega_c(1+z)^4}{\sqrt{(LH_0)^2E^2-1}}\,,
\ee
where now the dimensionless curvature radius becomes
\be
(LH_0)^2=1+\left(\frac{\Omega_c}{1-\Omega_m-\Omega_\Lambda}\right)^2\,,
\ee
and in limit ($\Omega_c\to0$, $L\to\infty$) we now recover the flat $\Lambda$CDM model. In this case the model contains two possible sources of low-redshift acceleration, and the interesting question is the extent to which data can distinguish between them.  Using the set $(\Omega_m, \Omega_\Lambda, LH_0)$ as independent parameters, the following low-redshift background cosmology data constraints were found
\bq
\Omega_\Lambda&=&0.70_{-0.03}^{+0.02}\\
\Omega_m&=&0.17_{-0.11}^{+0.08}\\
(LH_0)&>&1.02\,,
\eq
where again the latter is a two-sigma lower limit, and the model overfits the data. It is interesting to note that the standard preferred value of the cosmological constant is unchanged, and the effect of the additional component in the Friedmann equation is to decrease the matter density, with a non-zero contribution from $\Omega_c$.

\section{\label{extens}Extended datasets}

To obtain constraints on this class of models over the entire cosmic history, we use a modified version of the CLASS software \cite{Class1,Class2}, including the RT fluid which, for numerical convenience, we define as
\be
\Omega_{RT}(z)=\frac{\Omega_c(1+z)^4}{\sqrt{(LH_0)^2E^2-1}}\,.
\label{eq12}
\ee
When reporting the results of our analysis in the following section we will use $\Omega_{RT}\equiv\Omega_c/\sqrt{(LH_0)^2-1}$ without an argument to denote its present-day value. Note that this $\Omega_{RT}$ is analogous (but not identical) to the $\Omega_c$ used in the previous section. This redefinition addresses a numerical challenge: in the original formulation, the RT contribution depended explicitly on $H_0$, which, in the CLASS analysis, is only determined at the end of the integration. By redefining the fluid in terms of $\Omega_{RT}$, we decouple it from the evolving dimensionless Hubble parameter $E(z)$, allowing the entire RT contribution to be described by an effective equation of state
\be
p_{RT} = \frac{(L^2\ddot{a}/a-1)}{3 (L^2H^2-1)}\rho_{RT}\,. 
\label{State}
\ee
Based on the equation of state, the perturbation equations for $\delta_{RT}$ and $\theta_{RT}$ were implemented in CLASS and integrated consistently with the standard cosmological perturbations. The fluid's impact on background cosmology is computed by integrating the model equations to obtain $\rho_{RT}(z)$.

Our extended and updated cosmological data includes Planck 2018 constraints on CMB power-spectra and lensing \cite{Planck1,Planck2}, large scale structure and baryon acoustic oscillation data from the BOSS DR-12 galaxy survey \cite{Alam_2017}, as well as the Pantheon+ dataset \cite{Brout} and the Hubble measurements coming from a recent compilation of cosmic chronometer data \cite{Moresco}. We note that the latter three datasets are low-redshift ones, which are comparable (but not exactly identical) to the low-redshift datasets used in our earlier work \cite{artigo1}. In what follows we also refer to them as the low-redshift data, to be contrasted with the Planck data and the combination of all these datasets,  which we will refer to as 'All' in plots and tables.

The likelihood analysis is done by sampling Monte Carlo Markov Chains (MCMC) with MontePython \cite{Monte1,Monte2} directly coupled to the modified CLASS code. We consider the chains to be converged if, for all parameters, the Gelman-Rubin criterion satisfies $|R-1| < 0.05$.

For every run, we sample over the standard cosmological parameters ($\omega_b$, $\Omega_{cdm}$, $\Omega_\Lambda$, $H_0$), as well as the specific RT parameters ($\Omega_{RT}$, $L$) and the nuisance parameters of the various likelihoods. The priors in all of these parameters are flat and unbounded. The exception to this last point is $L$, which needs to be bounded. Specifically, one needs to constrain the lower values of $L$, to ensure that the square root present in the equation of state, Eq. (\ref{State}) is real (otherwise the corresponding parameters will evidently be unphysical). The upper limit also needs to be bound since $L$ can tend to infinity.

Bearing the above in mind, we have chosen a prior range $L\in[4900,20000]$ Mpc, corresponding to approximately 1 to 4 Hubble horizons. The lower prior originates from Eq. \ref{eq12}. For $z=0$, $E=1$, which trivially constrains the possible values of $L$, as $(LH_0)^2-1\geq 0$.  For the upper limit, 4 Hubble horizons was chosen because the data does not provide additional insights for distances beyond this scale. To confirm that the results regarding $L$ are not sensitive to the chosen upper limit, we have repeated the analysis extending the range to 10 Hubble horizons and obtained consistent results. Considering for example a redshift $z \approx 1$, for values significantly larger than 10 Hubble radii, the denominator of Eq. \ref{eq12} approaches $L H_0$, implying one of two possibilities: either $\Omega_c$ is sufficiently large for $\Omega_{RT}$ to remain relevant at $z = 1$, potentially causing issues at earlier times due to its $\approx(1+z)^4/LH(z)$ dependence, or $\Omega_c$ is small, making $\Omega_{RT}$ negligible. Since the data does not provide much insight for distances beyond this scale, values of $L$ much larger than 10 Hubble radii are of limited interest for study.

\section{Results}

\begin{figure*}
\begin{center}
\includegraphics[width=1.0\textwidth]{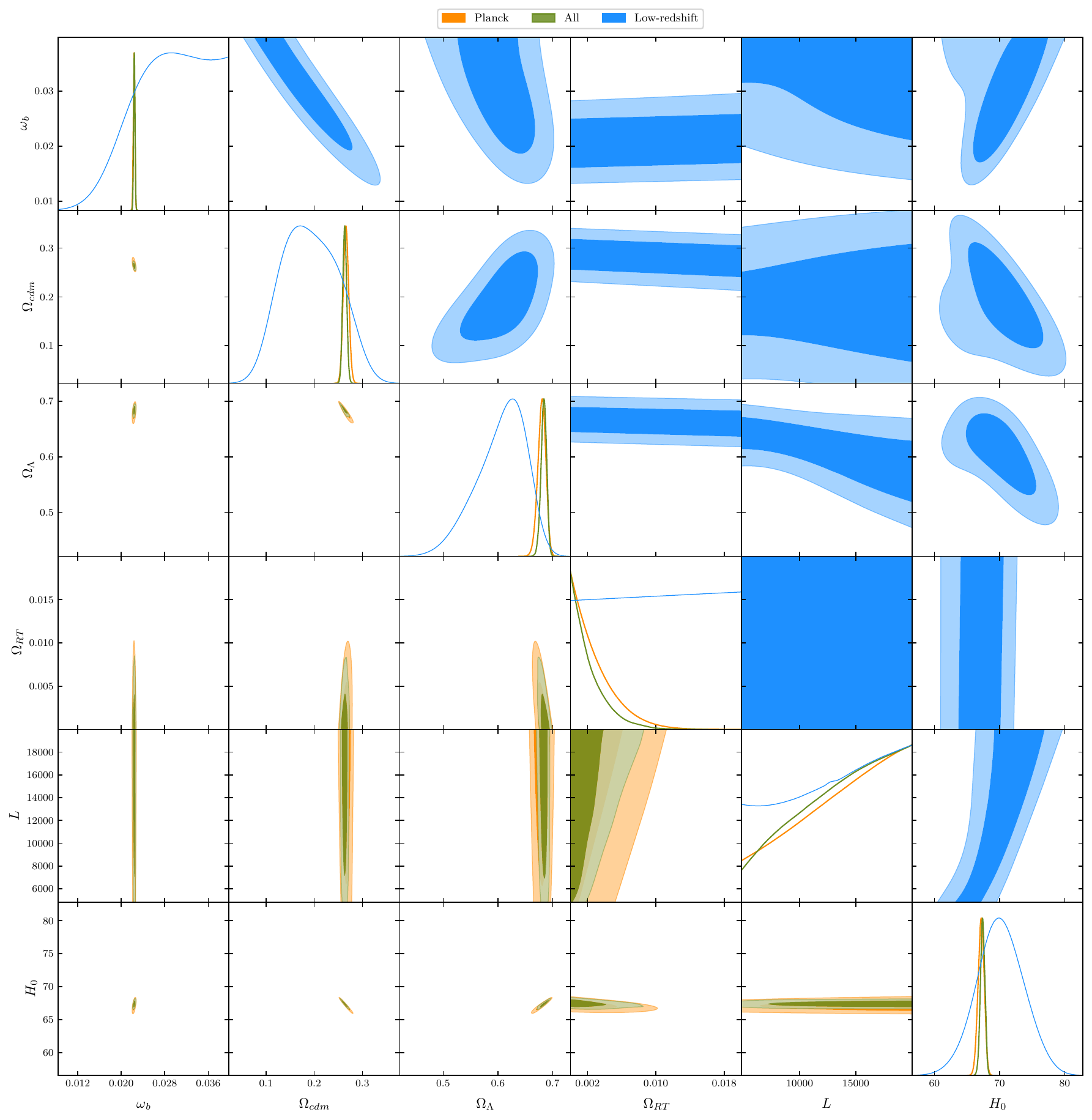}
\end{center}
\caption{Constraints on the Stern \& Xu realization of the Regge-Teitelboim model. Blue, orange and green correspond to the low-redshift, Planck and full datasets respectively. One and two-sigma confidence regions are shown in all two-dimensional panels, with the remaining ones showing the one-dimensional posterior likelihoods.}
\label{figure1}
\end{figure*}

\begin{figure*}
\begin{center}
\includegraphics[width=1.0\textwidth]{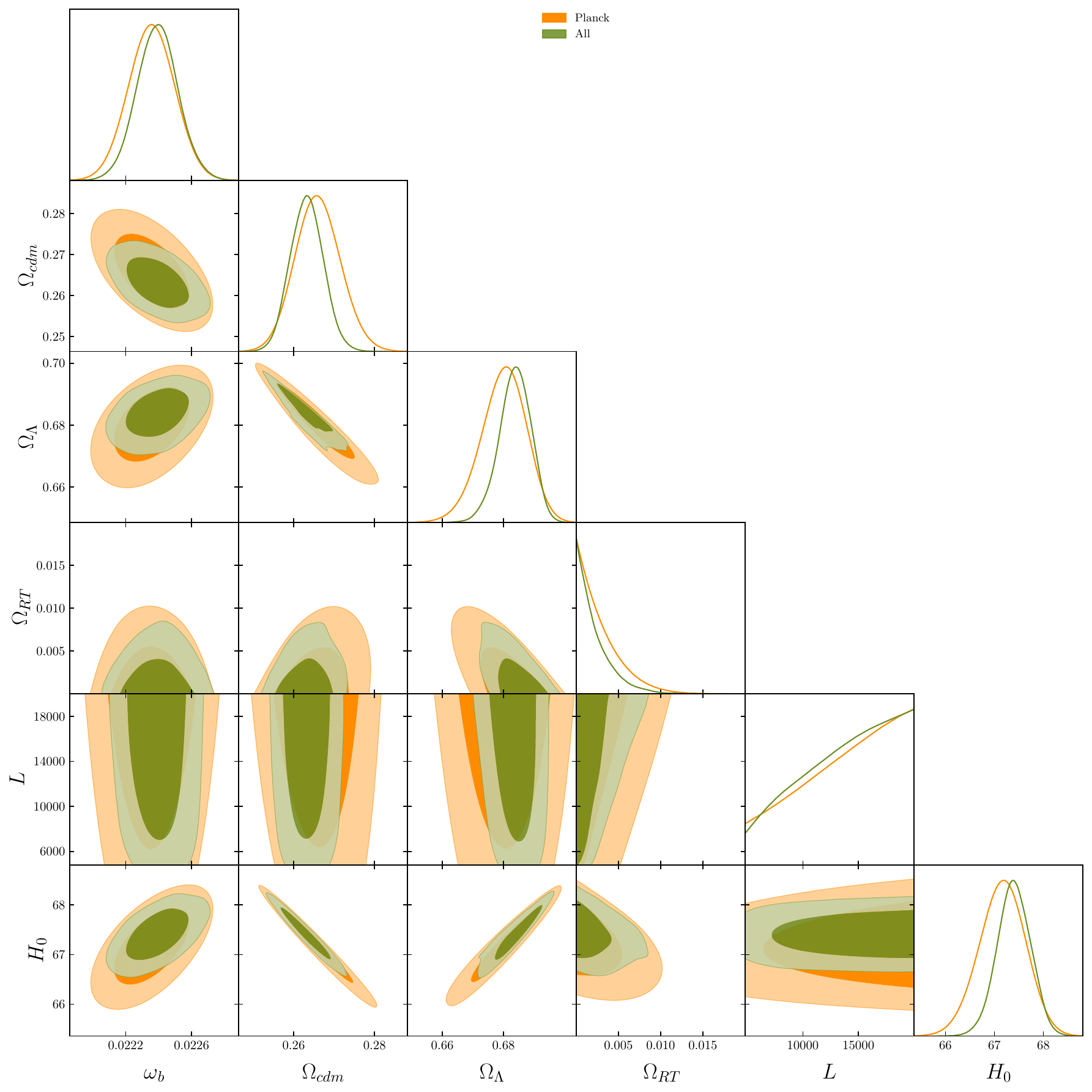}
\end{center}
\caption{Zoomed-in version of the the parameter space also shown in Figure \ref{figure2}, highlighting the differences in the constraining power of the Planck and full datasets.}
\label{figure2}
\end{figure*}

Table \ref{table1} and Figures \ref{figure1} and \ref{figure2} summarize the results of our analysis, listing the constraints on the relevant model parameters, separately for our low-redshift and Planck datasets described in the previous section, as well as for the combination of all the datasets described therein.

\begin{table}
\centering
\caption{One sigma posterior constraints, or two sigma limits, on the most relevant model parameter, for the three dataset choices detailed in the main text.}
\begin{tabular}{c | c | c | c}
\hline
Parameter & Low redshift & Planck & All \\
\hline
$H_0$ & $71.3_{-5.7}^{+3.6}$ & $67.2_{-0.4}^{+0.5}$ & $67.1\pm0.4$ \\
$\Omega_m$ & $0.25\pm0.09$ & $0.315\pm{0.012}$ & $0.313\pm0.009$ \\
$\Omega_\Lambda$ & $0.60_{-0.04}^{+0.07}$ &  $0.680_{-0.006}^{+0.008}$ & $0.679_{-0.006}^{+0.007}$ \\
$\Omega_{RT}$ & $0.22_{-0.14}^{+0.13}$ & $<0.007$ & $<0.006$ \\
$LH_0$ & $>1.15$ & $>1.43$ & $>1.45$ \\
\hline
\end{tabular}
\label{table1}
\end{table}


Starting with the constraints from low-redshift data, we see that these are broadly consistent from those in our earlier work, bearing in mind the different datasets and also the different parameters in the analysis. Specifically, in the current work there are more free parameters, e.g. in \cite{artigo1} the Hubble constant was analytically marginalized while in the present work is is a free parameter. The main consequence of this wider parameter space is that constraints on the model parameters become significantly weaker. Still, there is a mild preference for a significant contribution of $\Omega_{RT}$ to the universe's energy  budget.

The addition of the Planck and other lower redshift data provides extremely stringent constraints, effectively forcing the Regge-Teitelboim fluid density to be indistinguishable from zero, and the model to be confined to the close neighborhood of its $\Lambda$CDM limit. In other words, although the model contains two possible mechanisms for the low-redshift accelerations of the universe, Planck data can clearly distinguish between them, and the RT specific component is constrained to $\Omega_{RT}<0.006$ at the 95 percent confidence level. Similarly, the de Sitter curvature radius in units of the Hubble constant is constrained to $LH_0>1.45$.

\begin{figure*}
\begin{center}
\includegraphics[width=0.49\textwidth]{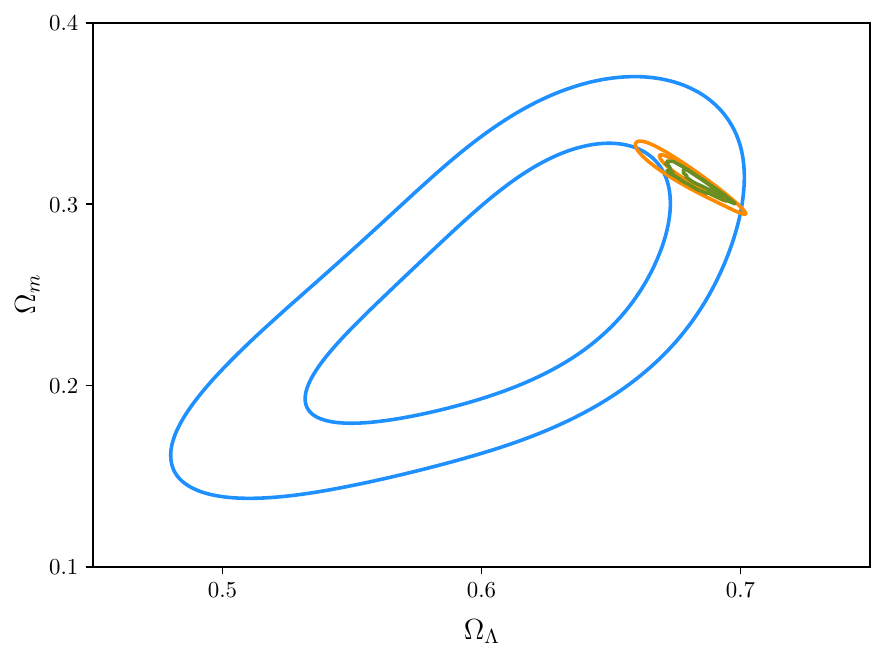}
\includegraphics[width=0.49\textwidth]{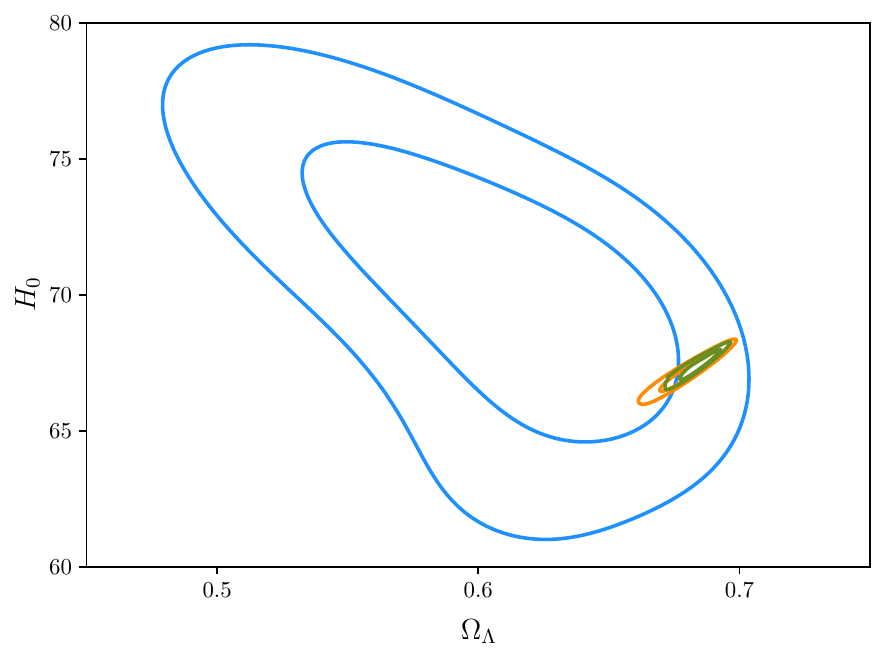}
\includegraphics[width=0.49\textwidth]{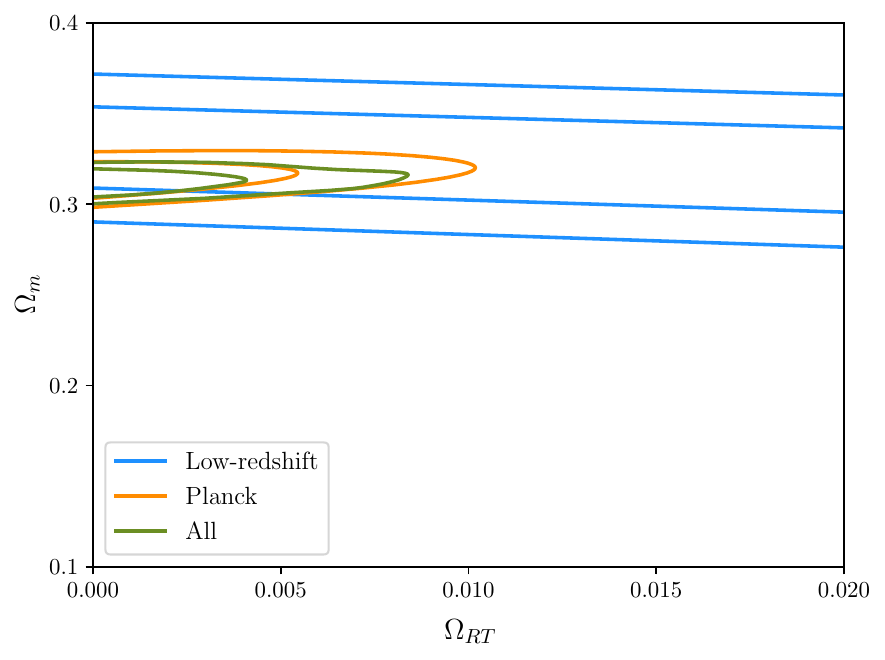}
\includegraphics[width=0.49\textwidth]{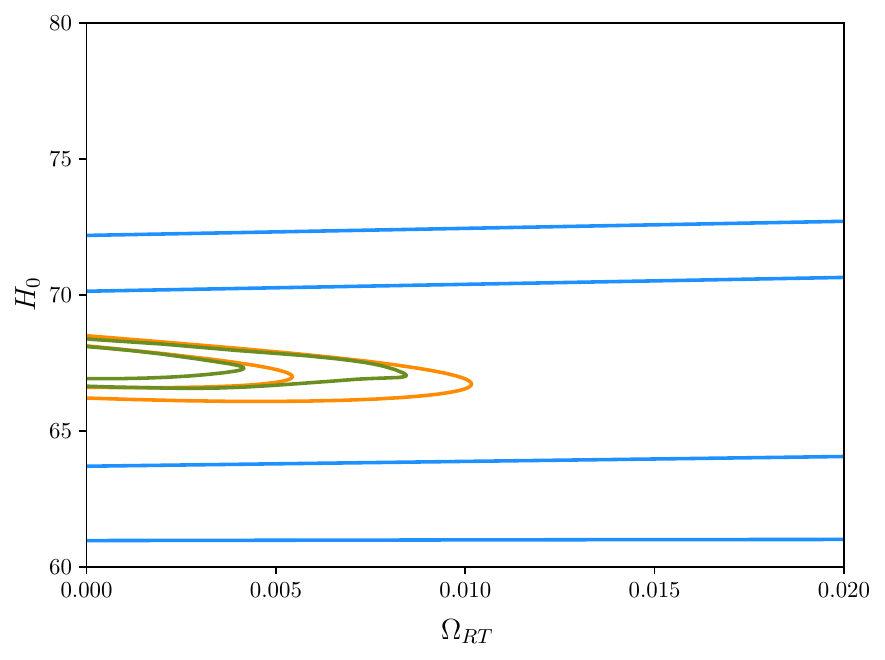}
\end{center}
\caption{Comparing constraints from the low-redshift, Planck and combined datasets in relevant two-dimensional parameter spaces, with the remaining parameters marginalized. One and two-sigma confidence regions are displayed throughout.}
\label{figure3}
\end{figure*}

Figure \ref{figure3} further highlights the differences in the constraining power of the various data sets for relevant two-dimensional planes. The overall conclusion is that the modified Stern \& Xu model is constrained, by Planck data, to be indistinguishable from the $\Lambda$CDM model. Here it's interesting to note that considering low redshift data alone $\Omega_m$ and $\Omega_\Lambda$ would be positively correlated, and both would negatively correlated with $\Omega_{RT}$. The Planck data effectively removes $\Omega_{RT}$ as a significant contributor, and consequently $\Omega_m$ and $\Omega_\Lambda$ become negatively correlated, as expected.

The bottom line is that there is no statistical preference for the RT model. This becomes quantitatively clear when applying the Akaike Information Criterion (AIC). The Stern \& Xu model has 8 fitted parameters, whereas $\Lambda$CDM has only 6. From the data, we calculate: $\Delta\text{AIC}= \text{AIC}_{\text{RT}}-\text{AIC}_{\Lambda\text{CDM}}=10.68$. According to widely accepted thresholds, $\Delta\text{AIC}>10$ is regarded as very strong evidence against the more complex model. This value clearly indicates that $\Lambda$CDM is strongly preferred by the data.

\section{Conclusions}
\label{concl}

We have built upon recent work on the Stern \& Xu class of Regge-Teitelboim gravity \cite{artigo1}, and provided updated and much tighter observational constraints on this model class. While the earlier work was restricted to low-redshift background cosmology data, in the present work we included additional data, and in particular the Planck CMB data, which is primarily behind the tightening of the constraints.

These models stem from the assumptions that gravity is described by embedding the usual space-time manifold in a fixed higher-dimensional background and that one can regard the embedding coordinates as dynamical degrees of freedom, which lead to specific source terms in the Einstein equations. The choice of embedding is effectively a further model degree of freedom, and the distinguishing feature of the Stern \& Xu is that it assumes a non-flat higher-dimensional background.

Phenomenologically, these models contain an acceleration mechanism alternative to the cosmological constant, which in principle is therefore not needed. However, the earlier work already showed that such a scenario was observationally unrealistic. The only plausibly realistic scenario for these models, therefore, is to consider the case where a cosmological constant is also allowed, making them parametric extensions of $\Lambda$CDM. In such case the model includes two possible acceleration mechanisms, and the interesting phenomenological question is therefore the extent to which observations can distinguish between them.

Our earlier work \cite{artigo1} has shown that low redshift data prefers a value of $\Omega_\Lambda$ which is consistent (within statistical uncertainties) with the standard one, with the additional embedding-related source term in the Friedmann equation having a non-negligible contribution and having the main effect of lowering the preferred value of the matter density. Our analysis in the present work, when confined to low redshift data, confirms this (despite the different assumptions on the model's free parameters, and on the treatment of the Hubble constant). 

However, the Planck CMB data tightly restricts the embedding term's contribution, as defined by $\Omega_{RT}$ to sub-percent level, and therefore forces the model to be extremely close to $\Lambda$CDM, which significantly limits the possible distinct phenomenological implications of this class of models. This is further reinforced by the AIC diagnostic, which quantitatively favors $\Lambda$CDM over the extended model. Phenomenologically, this suggests that embeddings of this class of models in non-flat higher-dimensional backgrounds carry no advantage as compared to flat ones. Our results thus further support the idea that $\Lambda$CDM, even if not fully correct, is a convenient and robust approximation to a more fundamental underlying model, still to be discovered.

\begin{acknowledgments}
This work was financed by Portuguese funds through FCT (Funda\c c\~ao para a Ci\^encia e a Tecnologia) in the framework of the project 2022.04048.PTDC (Phi in the Sky, DOI 10.54499/2022.04048.PTDC). CJM also acknowledges FCT and POCH/FSE (EC) support through Investigador FCT Contract 2021.01214.CEECIND/CP1658/CT0001 (DOI 10.54499/2021.01214.CEECIND/CP1658/CT0001).
\end{acknowledgments}

\appendix
\section{\label{appendix}The no baryonic matter limit}

Here we make a short digression to consider the constraints from the low-redshift data in our previous work \cite{artigo1} in the case where one has a cosmological constant but no matter density, i.e. setting $\Omega_m=0$.  (We thank Allen Stern for suggesting this scenario to the authors.) Phenomenologically, this artificial limit can be envisaged from the difference between baryonic and no-baryonic matter. In the RT approach, $\Omega_m$ might be interpreted as referring only to the visible baryons, while the 'traditional' non-baryonic matter would be replaced and accounted for by $\Omega_c$. Thus the assumption of $\Omega_m=0$ crudely corresponds to the simplifying assumption that baryons are unimportant for the cosmological dynamics. Moreover, this exercise is useful for highlighting the interplay between the model parameters $\Omega_c$ and $\Omega_\Lambda$---or, in other words, between the two possible mechanisms leading to acceleration.

\begin{figure*}
\begin{center}
\includegraphics[width=0.49\textwidth]{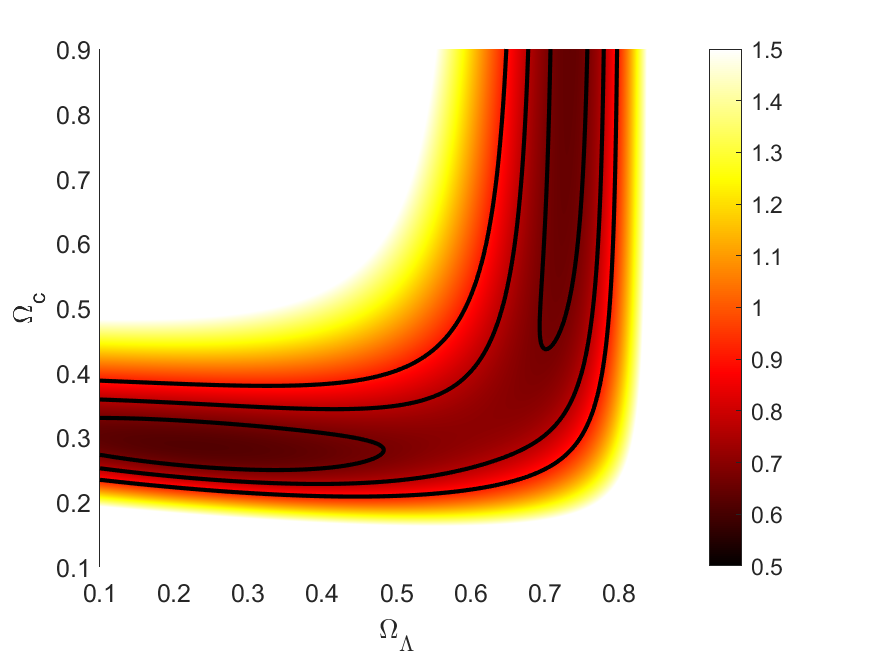}
\includegraphics[width=0.49\textwidth]{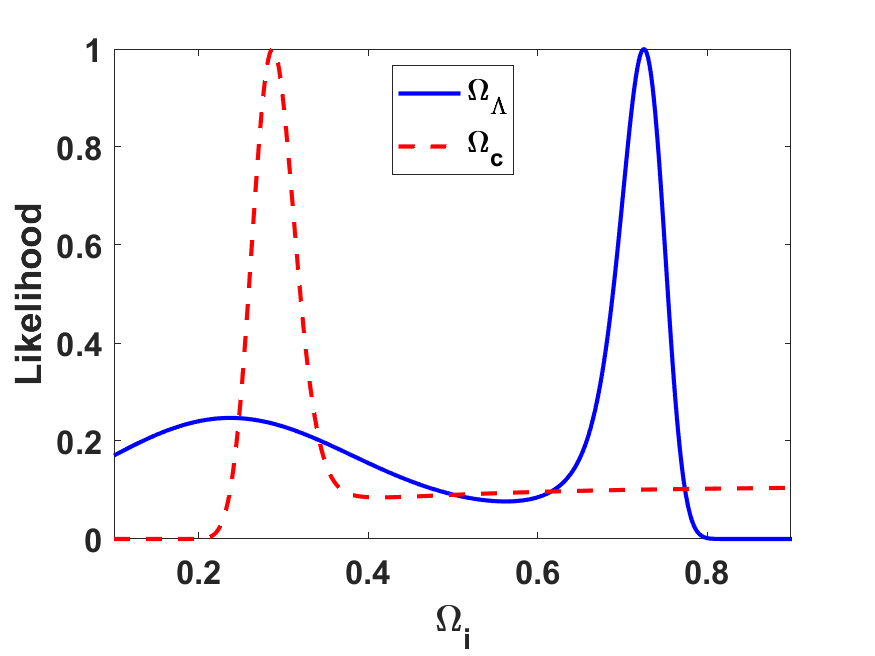}
\end{center}
\caption{Low-redshift constraints on the the Stern-Xu class of Regge-Teitelboim models including a cosmological constant but no baryonic matter. In the left panel the solid black lines show the one, two and three sigma constraints in the $\Omega_c$--$\Omega_\Lambda$, while the color map shows the reduced chi-square, for a total of 42 degree os freedom. The right panel shows the one-dimensional posterior likelihoods for each of the parameters, with the other marginalized.}
\label{figureA}
\end{figure*}

In this case we are left with a two-dimensional parameter space $(\Omega_c, \Omega_\Lambda)$, and figure \ref{figureA} shows the results of the analysis, which leads to the constraints
\bq
\Omega_\Lambda&=&0.73_{-0.03}^{+0.02}\\
\Omega_c&=&0.29_{-0.02}^{+0.03}\,.
\eq
While the two parameters are partially degenerate, both can be reasonably constrained and once again the preferred value of the cosmological constant is consistent with the standard one, while the additional higher-dimensional induced term in the Friedmann equation, parameterized by $\Omega_c$, takes over the role of the matter density. This case therefore highlights the interplay between the two parameters in this model.

\bibliography{paper}
\end{document}